# Operando monitoring of strain field distribution in lithium battery anode via ultra-high spatial resolution optical frequency domain reflectometer

Kaijun Liu[1]‡, Zhijuan Zou[3]‡, Guolu Yin[1,2]*, Yingze Song[3]*, Zeheng Zhang[1], Yuyang Lou[1], Zixuan Zhong[1], Huafeng Lu[1], Duidui Li[1], and Tao Zhu[1,2].

1 The Key Laboratory of Optoelectronic Technology and Systems (Ministry of Education), Chongqing University, Chongqing 400044, China

2 State Key Laboratory of Coal Mine Disaster Dynamics and Control, Chongqing University, Chongqing 400044, China

3 State Key Laboratory of Environment-FriendlyEnergy Materials, School of Materials and Chemistry, Center ofAnalysis and Charac-terization, Tianfu Institute of Researchand Innovation, Southwest University of Science andTechnology, Mianyang 621010, China

**ABSTRACT:** The cycling performance of lithium-ion batteries is closely related to the expansion effect of anode materials during charge and discharge processes. Studying the mechanical field evolution of anode materials is crucial for evaluating battery performance. Here, we propose a phase-sensitive ultra-high spatial resolution optical frequency domain reflectometry technique, in which the test fiber is embedded into the anode of a lithium-ion battery to monitor the mechanical evolution of the anode material during cycling. We investigated the strain evolution of the anode material under different loading levels and used this method to infer the morphological changes of the material. Furthermore, combining this with battery capacity information provides a new approach for assessing the performance of lithium-ion batteries.

## INTRODUCTION

In recent years, with the rapid development in the field of new energy vehicles, higher requirements have been proposed for lithium-ion batteries, specifically in terms of high energy density and high stability.[1] Therefore, the selection of anode materials for lithium-ion batteries is crucial. Graphite is a commonly used anode active material for lithium-ion batteries, but it has a low theoretical capacity.[2-3] As a result, researchers have shifted their focus to Si-based materials.[4] Pure silicon has a theoretical capacity 10 times higher than that of graphite, but its significant volume expansion effect leads to rapid degradation of cycling performance, leaving considerable room for improvement before practical application.[5-7] Silicon monoxide is an effective alternative with a smaller volume expansion effect relative to pure silicon, resulting in more stable cycling performance.[8-9] However, this is still insufficient for standalone use as an anode material. Therefore, incorporating it into graphite to form a graphite/SiO composite anode is an effective strategy.[10-11] The internal strain of the composite anode remains a major factor influencing its cycling performance. Currently, the coupling regulation mechanism between the electrochemistry of the battery and the mechanical deformation of the materials is still not well understood and lacks quantitative analysis.

During battery cycling, electrode materials undergo periodic strain, leading to stress accumulation within the battery due to the formation of dead lithium and solid electrolyte interphase (SEI). This results in deformation and cracking of electrode materials, thus deteriorating battery performance. Traditional approaches rely on strain gauges or pressure sensors to measure battery thickness,[12-13] but achieving in-situ measurements is challenging. Optical imaging techniques such as optical ranging can measure surface strain on the battery,[14] but experimental setups are complex and highly influenced by the environment, limiting measurements to laboratory settings. Additionally, schemes like X-ray photoelectron spectroscopy can measure internal battery strain but require battery disassembly.[15] Furthermore, as they are based on non-contact measurements, their measurement accuracy is typically low. Therefore, the development of an advanced in-situ measurement technique for batteries is crucial.

Due to their small size, optical fiber sensors can be easily integrated into the interior of batteries, and their resistance to electromagnetic interference ensures that they do not interfere with the normal operation of the battery.[16] Moreover, their stable physical and chemical properties enable them to operate inside batteries for extended periods. Fiber Bragg gratings (FBGs) are widely used for monitoring the strain or temperature in the batteries. Guo et al. utilized FBGs to monitor thermal runaway within lithium-ion batteries,[17] and tilted fiber Bragg gratings were employed to observe dendrite formation in lithium metal batteries.[18] Furthermore, Tarascon et al. used FBGs to monitor stress variations in the silicon cathode of button batteries,[19] while Huang et al. further investigated the impact of different diameters of silicon particles on battery performance.[20] However, the length of an FBG is typically around 1 cm, making it essentially a point sensor with measurement results reflecting the average strain within a 1 cm spatial range. Given that the length and width of pouch-type lithium-ion battery negative electrode materials are approximately 4-6 cm, and uneven coating is possible during cathode material application, relying solely on strain measurements from a single point may not accurately reflect overall battery performance.

To address this limitation, we propose a phase-sensitive high spatial resolution optical frequency domain reflectometry (φ-OFDR) technique for distributed strain evolution monitoring of graphite/SiO composite anodes in lithium-ion batteries. We conducted simulations and experiments to study the high spatial resolution and detailed resolution capability of φ-OFDR, demonstrating its strain evolution results at different positions within the negative electrode. We characterized the strain evolution of active materials under different levels of loading, revealing the relationship between the content of active materials and strain. Finally, by combining strain with various electrochemical parameters, we analyzed their impact on battery performance.

## EXPERIMENTAL METHODS

**Materials and Electrode Preparation.** In this experiment, all silicon-based active materials were composed of silicon monoxide (SiO)/graphite composite materials (theoretical specific capacity of 450 mAh/g). The SiO/graphite composite materials, conductive carbon black (CB), carboxymethyl cellulose (CMC, 1.5 wt%), and styrene-butadiene rubber (SBR, 5 wt%) were thoroughly mixed in a mass ratio of 8:1:0.5:0.5 using a planetary mixer to prepare the slurry. The resulting slurry was uniformly coated onto copper foil and dried in a 60°C vacuum oven for 12 hours. Three different loading amounts of the anode were prepared, namely 1.11 mg/cm$^2$, 1.52 mg/cm$^2$, and 2.00 mg/cm$^2$.

**Pouch Cell Assembly.** (1) The pre-prepared silicon-based negative electrode was cut to a specific size (4*6 cm). Aluminum tabs with earplugs were spot-welded onto the silicon negative electrode using a spot welding machine.

(2) A lithium foil (thickness: 80 μm) was adhered onto the copper foil, slightly larger than the silicon negative electrode. Nickel tabs with earplugs were spot-welded onto the copper-lithium composite positive electrode using a spot welding machine.

(3) For the optical fiber sensing analysis of the silicon negative electrode, single-mode optical fibers were arranged and secured on the negative electrode surface in accordance with ...

(4) The silicon negative electrode with arranged single-mode optical fibers, separator (2500 Monolayer Microporous Membrane, Celgard), and copper-lithium composite positive electrode were stacked from bottom to top. The separator served to separate the positive and negative electrodes.

(5) Rubber tabs were adhered to the heat-sealing position of the optical fibers to secure them onto the aluminum-plastic mold and prevent electrolyte leakage. After filling with 1 mL of lithium electrolyte, the soft pack battery was allowed to stand for 12 hours before testing. The lithium electrolyte was purchased from Coulomb and contained 1.0 M LiPF$_6$, with an EC:DEC:EMC ratio of 1:1:1 vol%. All operations were conducted inside a glove box filled with argon (H$_2$O < 0.01 ppm, O$_2$ < 0.01 ppm).

**Electrochemical Tests.** To test the battery performance, the assembled lithium battery is placed in a constant temperature test chamber (JOINTEC, SPX-150BIII) at 30 °C, and the positive and negative electrodes of the battery are connected to the battery testing system (NEWARE, CT-4008Tn-5V6A-S1-F) for charge-discharge testing. The battery operates at a charging rate of 0.5 C after being activated with a 0.1 C charge in the first cycle, with a voltage range of 0.01-2 V.

**System Design and Sensing Principle of OFDR.** The OFDR system we employed is illustrated in Figure 1(a). The frequency-swept laser emitted by the swept laser source is split into two paths. One path enters the Mach-Zehnder interferometer, serving as the main interferometer located below. The scattered light from the test fiber passes through a circulator and interferes with the reference light to generate a beat frequency signal. The beat frequency is proportional to the distance information of the fiber, thus achieving positioning. The auxiliary interferometer above is used for non-linear correction of the frequency-swept light, suppressing spectral broadening to enhance the system's spatial resolution. The optical signal is converted into an electrical signal through a balanced detector and collected by a data acquisition card. We utilized a single-mode optical fiber coated with polyimide as the sensing unit, and encapsulated it through a serpentine routing method. The optical fiber was subjected to certain pre-stress and fixed at both ends, ensuring close attachment to the anode material, as shown in Figure 1(b). When the anode material undergoes expansion, it will stretch the optical fiber, generating axial strain, which is demodulated by the OFDR system. The demodulation principle is depicted in Figure 1(c), where the optical fiber contains numerous scattering points, each with distinct fingerprint spectra. When the test fiber is stretched or compressed,[21] changes in length and refractive index due to the Poisson and elasto-optic effects cause phase variations:

$$\Delta\varphi_z = diff(\Delta\varphi) = \frac{4\pi}{\lambda_0}\left[(1-P_e)n_{eff}dz\right], \quad (1)$$

Here, $diff$ represents the numerical differentiation symbol, $\lambda_0$ is the laser wavelength, $P_e$ is the elasto-optic coefficient, $n_{eff}$ is the effective refractive index, and $dz$ is the length change at the position $z$ of the optical fiber. This causes a drift in the local fingerprint spectrum at the position of the scattering point, which enables strain demodulation by calculating the wavelength drift. However, this approach has limitations in sensor spatial resolution. To ensure microstrain-level demodulation accuracy, the sensor spatial resolution is typically in the range of 2-4 cm, which significantly limits its application for distributed measurements inside batteries. As the essence of strain in OFDR is the phase variation, it can be expressed as:

$$\varepsilon = \frac{dz}{\Delta z} = \frac{\lambda_0 \Delta\varphi_z}{4\pi(1-P_e)n_{eff}\Delta z}, \quad (2)$$

Here, $\Delta z = \lambda_0^2/(2n_{eff}\Delta\lambda)$ represents the spatial resolution, where $\Delta\lambda$ is the wavelength range of the swept source.[22] Therefore, we adopted the φ-OFDR, which has the potential to achieve higher sensing resolution. In our previous work, we have demonstrated its capability to achieve a sensing spatial resolution of ~1.5 mm and a sensing accuracy at the microstrain level (~1 με).[21]

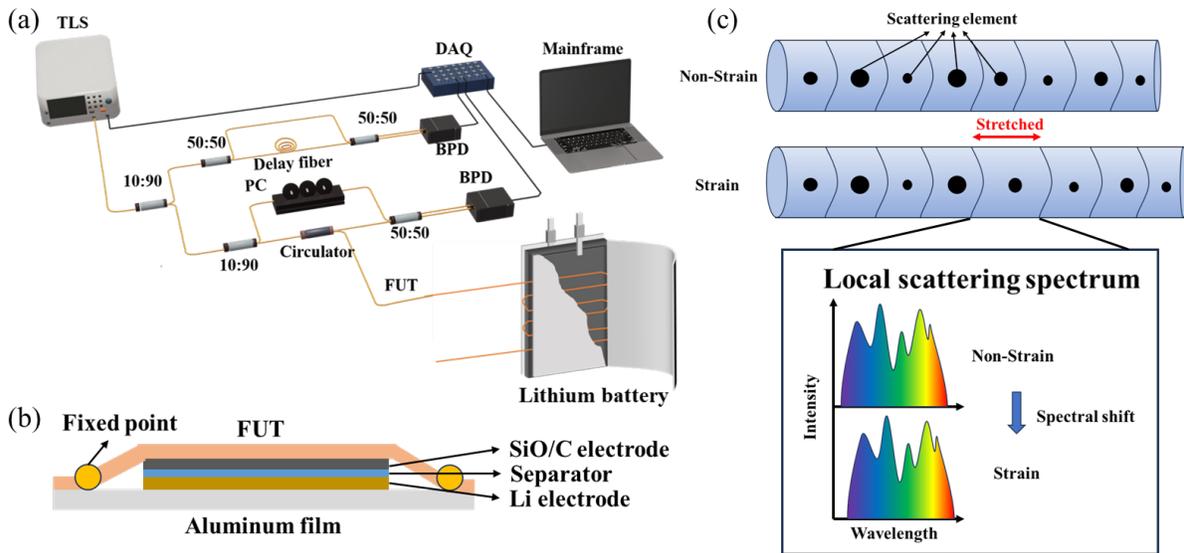

Figure 1. (a) The OFDR system setup for battery in-situ monitoring. (b) Schematic diagram of optical fiber/battery packaging structure. (c) Strain schematic diagram of optical fiber scattering element; Illustration: local scattering spectrum shift

## RESULTS AND DISCUSSION

**Sensing response characteristics of φ-OFDR.** To demonstrate the high sensing spatial resolution capability of φ-OFDR in monitoring battery anodes, we conducted a simple simulation analysis. We established a model as shown in the upper figure of Figure 2(a), creating a three-layer model: the bottom layer represents the anode material, the middle layer represents the optical fiber fixed at both ends, and the top layer represents the pouch cell casing. We simulated the expansion effect by applying upward stress to the anode material. The axial strain distribution of the battery is depicted in Figure 2(b). Due to the inability to strictly correspond the material parameters in the simulation to real-world values, we normalized the simulation results. It can be observed that the optical fiber on the anode material shows a uniformly distributed strain, with a depressed strain distribution characteristic from the fixed points of the optical fiber to the middle area of the anode material. We compared the traditional OFDR correlation demodulation results with the phase demodulation results as shown in Figure 2(c). It is evident that the phase demodulation results exhibit the characteristic of the depressed distribution, which is not effectively reflected in the correlation demodulation results. Furthermore, the phase demodulation results show clearer detailed information. We reconstructed the strain distribution of the anode from the measurement results of the serpentine routing encapsulated structure, as shown in Figure 2(d). A clear strain distribution field is presented within a spatial range of 4*6 cm, corresponding to the size of the anode we coated.

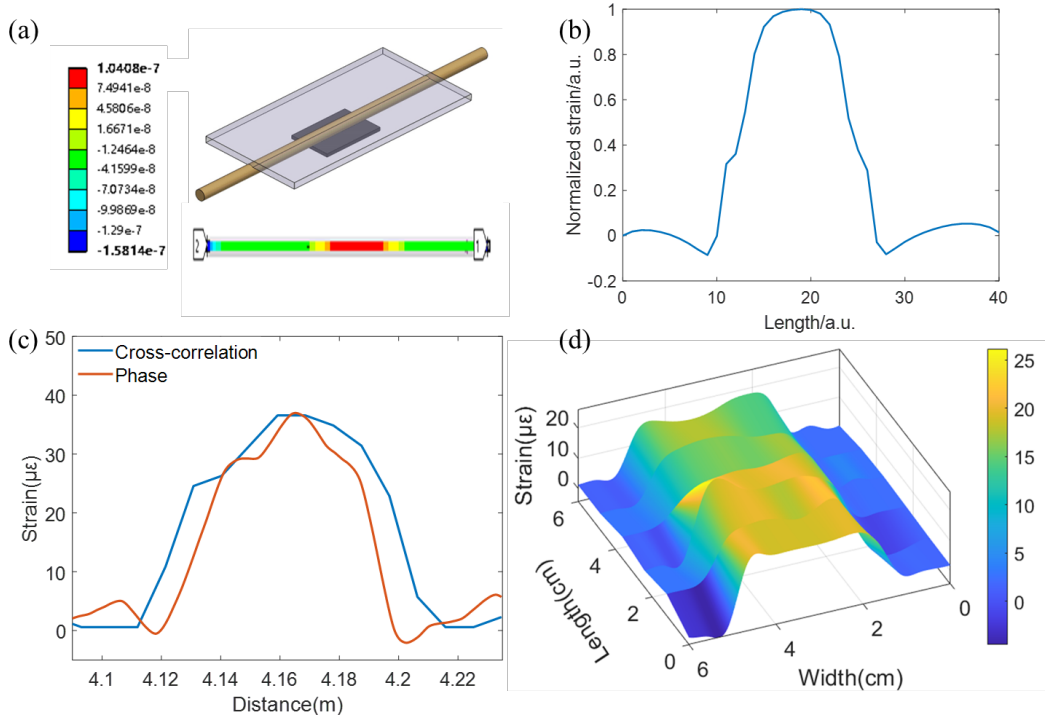

Figure 2. (a) ANSYS FEA simulation for optical fiber strain under the expansion of anode material. (b) Simulation results of optical fiber axial strain. (c) Comparison of cross-correlation demodulation and phase demodulation algorithms. (d) Demodulation results of two-dimensional strain field based on φ-OFDR.

**Cyclic response characteristics of battery.** To demonstrate the effectiveness of distributed measurements, we randomly selected 6 points along the distributed optical fiber and labeled them with different colors, as shown in Figure 3(a). The strain evolution patterns are shown in Figure 3(b-g). All the location points exhibited similar cyclic characteristics, with some differences possibly arising from the unevenness of the material coating shown in Figure S1.

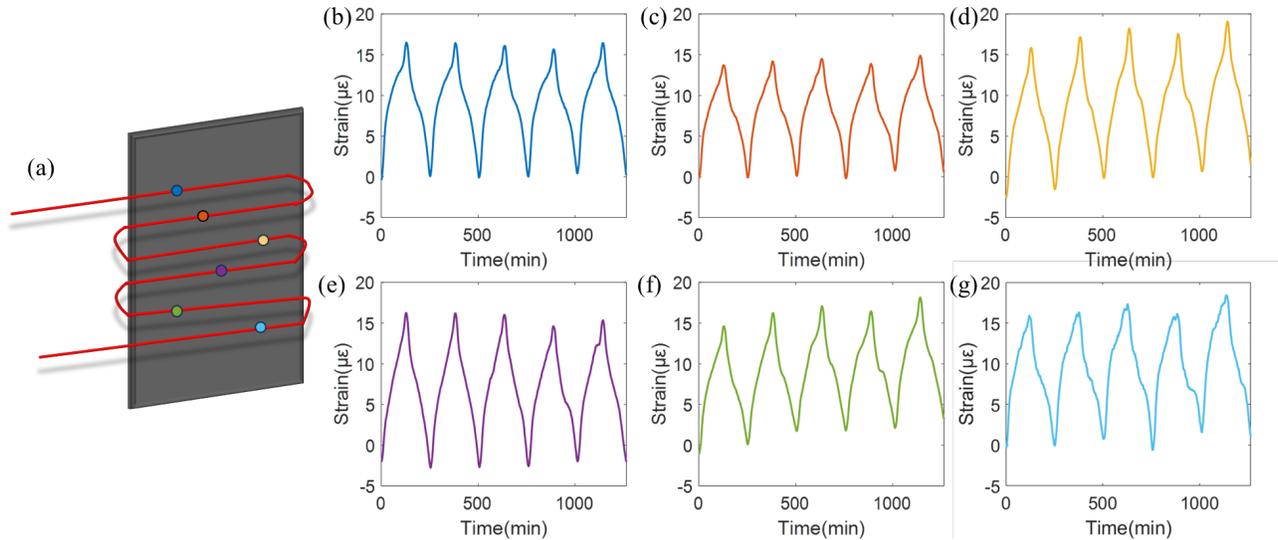

Figure 3. Evolution of anode strain field. (a) Schematic diagram of different measuring positions. (b-g) Evolution of strain with time at different positions.

In order to better analyze the performance of the battery, we correlated the strain measurements with the electrochemical parameters. Figure 4(a) compares the voltage and average strain as a function of time. The cyclic characteristics of the battery voltage are found to be in strict consistency with the strain evolution pattern. During the charging process, the strain shows a decreasing trend, corresponding to lithium extraction, while during the discharging process, the strain exhibits an increasing trend, correlating to lithium insertion. Figure S2 represent scanning electron microscope images of the anode material thickness direction, illustrating the variation in anode thickness during battery charging and discharging processes. It can be observed that as the SOC decreases during discharge, an increase in anode material thickness causes axial tension in the fibers, thereby increasing strain. Conversely, during charging, an increase in SOC leads to a decrease in anode material thickness, causing the fibers to return to their unstretched state, resulting in decreased strain, consistent with theoretical and experimental results. Typically, it is difficult to directly obtain the SOC of a battery. To better analyze this parameter, we established a relationship model between strain and SOC as shown in Figure 4(b). During the discharging process, it is evident that as the SOC decreases, the strain first undergoes a slow and flat phase. A clear inflection point appears at 75%, after which the strain rapidly increases. At around 5% SOC, the strain exhibits a slow change again (~18 με), and then rapidly increases to its maximum value (20 με) near 0% SOC. In the charging process, the strain shows a uniform and monotonic change, decreasing correspondingly as the SOC increases. This characteristic curve can effectively help us predict the SOC of the battery based on the magnitude of strain in the anode material. Additionally, we also demonstrated the relationship model between strain and voltage as shown in Figure 4(c). This graph provides a clear conclusion: when the voltage exceeds 0.6V, the anode material essentially does not expand or contract. Expansion deformation predominantly occurs in the low-voltage range. This is due to the lower lithium (Li) extraction/insertion potential of the SiO/C composite anode. When the potential is low, Li ions more easily migrate from the positive electrode to the negative electrode and vice versa during charging. This implies higher discharging/charging efficiency of the battery. We further established a strain rate model and compared it with the charge change rate, showing a high degree of consistency, as depicted in Figure 4(d). Both curves exhibit three distinct peaks at 0.18V, 0.08V, and 0.025V, indicating rapid strain changes at these voltages. The two cathodic peaks at 0.025 V and 0.18 V correspond to the alloying processes of silicon, while the other cathodic peak at 0.08 V represents the alloying process of graphite.[23]

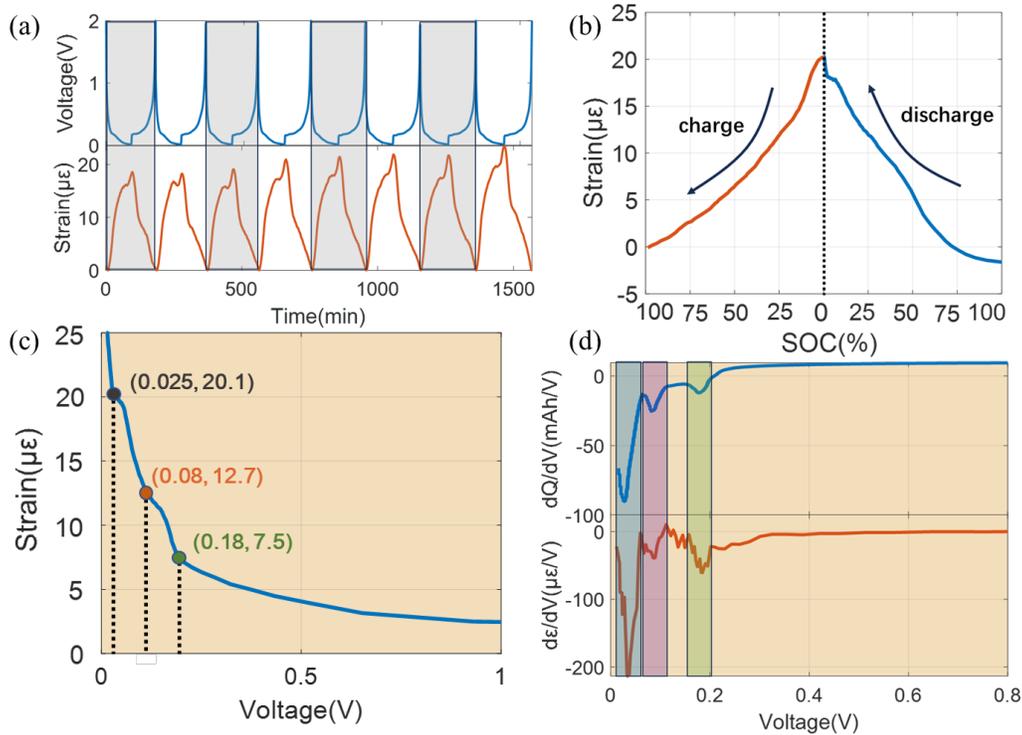

Figure 4. (a) Comparison diagram of voltage and strain variation with time. (b) Variation of strain with SOC during charge and discharge processes. (c) Relationship between strain and voltage during discharge process. (d) Comparison graph of charge and strain rate variation during discharge process.

The content of active materials in the electrodes determines the capacity and energy density that a battery pack can deliver, with a thicker electrode design implying higher energy output potential. However, continuously increasing the content of active materials can also lead to some adverse effects. Higher active material content will result in more lithium ions binding with the SiO/C, leading to a larger expansion effect and making it prone to positive electrode cracking and detachment during drying processes. Therefore, the strain evolution at three different loading levels of 1.11 mg/cm$^2$, 1.52 mg/cm$^2$, and 2.00 mg/cm$^2$ was investigated. We utilized strain-SOC curves for easy comparison, revealing that as the loading level increases, the maximum peak strains are 7.5, 14, and 20 με, showing an increasing trend. Hence, when designing batteries, a balance between capacity and stability should be considered by choosing an appropriate content of active materials.

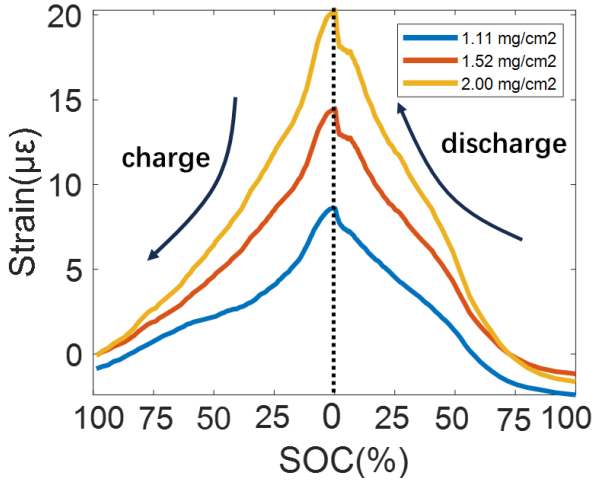

Figure 5. Variation of strain with SOC under different loadings of anode materials

Finally, we investigated the strain evolution characteristics over the entire battery cycle life. During the lithium insertion process, the crystal structures of carbon and silicon undergo expansion. The carbon particles limit the external expansion of the silicon particles, maintaining the stability of the electrode structure and the solid electrolyte interface film to prevent electrode detachment. During lithium extraction, the crystal structures of carbon and silicon undergo a contraction process; However, due to the presence of residual lithium, they do not fully revert to their initial geometric shapes. In other words, when lithium is repeatedly embedded/extracted from the electrode material, this process leads to periodic expansion/contraction of the active particle microcrystals. Accumulation of residual tensile strain occurs at the end of each cycle, resulting in corresponding tensile stress at each stage, which may accelerate the fracture of the brittle SiO@C composite electrode.

By comparing the strain magnitudes at different stages, we observed that in the first stage, there were relatively large fluctuations in strain at the initial phase, with a noticeable increasing trend. This may be attributed to the initial battery activation, and correspondingly, there was a significant increase in battery capacity during this stage. The slight increase in capacity during the initial cycles of lithium-ion batteries can be attributed to several factors. Firstly, the formation of a solid electrolyte interface during the first charge-discharge cycles increases the chemical reaction surface area and enhances charge transfer efficiency. Secondly, the process of lithium ion insertion into the negative electrode material leads to an increased uptake of lithium ions, resulting in higher battery capacity. Additionally, the gradual infiltration of the electrolyte into the electrode materials improves their conductivity, thereby enhancing lithium ion transport and contributing to the observed increase in battery capacity during the early cycles. It is important to note that this initial capacity increase is temporary and does not reflect a permanent expansion of the battery's actual capacity. With continued cycling, the battery capacity will stabilize and approach its rated value. In the second stage, the overall stress curve exhibited minor fluctuations, corresponding to a more stable battery performance phase. In the third stage, the maximum strain of the cycling period began to decrease significantly, and a noticeable strain accumulation appeared at the 135th cycle, which was caused by the presence of dead lithium and some lithium staying on the anode as part of the solid electrolyte interface (SEI). We analyzed the trends in both the overall cycling period capacity and the maximum strain, and found that the trend of strain changes was consistent with that of the capacity. A clear decrease in strain trend was evident around the 93rd cycle in the strain change curve, while in the capacity curve, this inflection point occurred at approximately the 100th cycle. This can effectively help us provide an early warning of battery failure. Therefore, we can conclude that the effective monitoring of strain can reflect the health status of the battery.

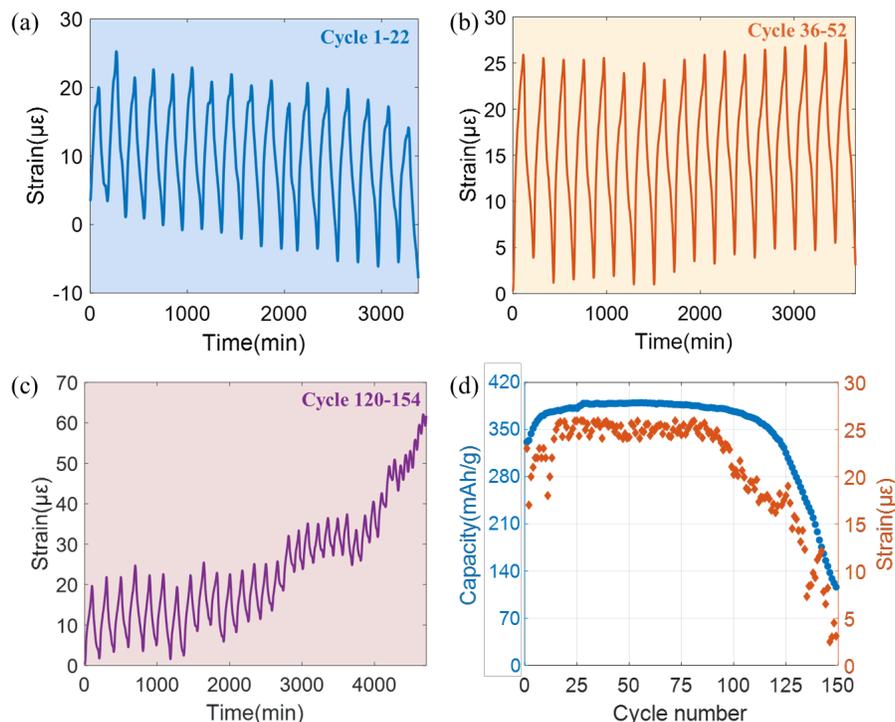

Figure 6. Strain evolution during full-life cycle of the battery. (a) Battery activation during early cycling stages. (b) Stable battery performance during mid-cycling stages. (c) Stress stacking during late cycling stages. (d) Evolution trend of battery capacity and maximum strain during the cycling process.

## CONCLUSIONS

In this work, we have demonstrated the application of φ-OFDR with mm-level spatial resolution in the field of in-situ strain field monitoring of lithium battery anode materials. The study investigates the temporal variation of strain in battery anodes, establishes the relationship between strain and battery SOC, and due to the abundant high spatiotemporal resolution data of φ-OFDR, it has the potential to predict battery SOC. By studying the evolution of strain and strain rate with voltage, the detachment and insertion patterns of anode materials and Li ions can be effectively analyzed. Additionally, the strain variation under different loadings of anode materials was analyzed, indicating that with increasing loading, the maximum strain of the anode during cycling increases, increasing the likelihood of anode material cracking, providing effective theoretical support for selecting appropriate anode material loadings. Finally, an analysis of the entire battery cycling process indicates that the maximum strain during the cycling period can effectively reflect battery life, and the inflection point of maximum strain precedes the capacity inflection point by ~7 cycles, which can help us anticipate battery life in advance.

## ASSOCIATED CONTENT

### Supporting Information

The Supporting Information is available free of charge on the ACS Publications website.

SEM; Surface image of SIO/C mixed anode; Thickness variation of anode under different SOC (PDF)

## AUTHOR INFORMATION

### Corresponding Author

**Guolu Yin** − The Key Laboratory of Optoelectronic Technology and Systems (Ministry of Education), Chongqing University, Chongqing 400044, China;
Email: glyin@cqu.edu.cn.

**Yingze Song** − State Key Laboratory of Environment-Friendly-Energy Materials, School of Materials and Chemistry, Center ofAnalysis and Characterization, Tianfu Institute of Researchand Innovation, Southwest University of Science andTechnology, Mianyang 621010, China;
Email: yzsong@swust.edu.cn.

**Author Contributions**

**Kaijun Liu** − The Key Laboratory of Optoelectronic Technology and Systems (Ministry of Education), Chongqing University, Chongqing 400044, China.

**Zhijuan Zou** − State Key Laboratory of Environment-Friendly-Energy Materials, School of Materials and Chemistry, Center of Analysis and Charac-terization, Tianfu Institute of Research and Innovation, Southwest University of Science and Technology, Mianyang 621010, China.

**Zeheng Zhang** − The Key Laboratory of Optoelectronic Technology and Systems (Ministry of Education), Chongqing University, Chongqing 400044, China.

**Yuyang Lou** − The Key Laboratory of Optoelectronic Technology and Systems (Ministry of Education), Chongqing University, Chongqing 400044, China.

**Zixuan Zhong** − The Key Laboratory of Optoelectronic Technology and Systems (Ministry of Education), Chongqing University, Chongqing 400044, China.

**Huafeng Lu** − The Key Laboratory of Optoelectronic Technology and Systems (Ministry of Education), Chongqing University, Chongqing 400044, China.

**Duidui Li** − The Key Laboratory of Optoelectronic Technology and Systems (Ministry of Education), Chongqing University, Chongqing 400044, China.

**Tao Zhu** − The Key Laboratory of Optoelectronic Technology and Systems (Ministry of Education), Chongqing University, Chongqing 400044, China.

**Author Contributions**

‡K. L. and ‡Z. J. Z. contributed equally to this work. The manuscript was written through contributions of all authors. All authors have given approval to the final version of the manuscript. K. L.: data curation, and writing the original draft. Z. J. Z.: Battery preparation and packaging. Z. Z., Y. L. and Z. X. Z.: data curation and formal analysis. H. L. and D. L.: formal analysis. G. Y., Y. S. and T. Z.: concept ualization, supervision, funding acquisition, resources, writing, review, and editing.

**Notes**

The authors declare no competing financial interest.

## ACKNOWLEDGMENT

This research was supported in part by the National Key Research and Development Program of China (2023YFF0715700), in part by the Fundamental Research Funds for the Central Universities (2023CDJKYJH064), in part by the Chongqing Talents: Exceptional Young Talents Project (CQYC202005011)